\documentstyle{elsart}
\begin{document}
\begin{frontmatter}
\title{The Proton-Proton Reaction, Solar Neutrinos, and a 
Relativistic
Field Theoretic Model of the Deuteron}

\author{John N. Bahcall\thanksref{jnb}}
\address{School of Natural Sciences, Institute for Advanced Study,
Princeton, NJ 08540}
\author{Marc Kamionkowski\thanksref{kamion}}
\address{Department of Physics, Columbia University, 538 West
120th Street, New York, New York~~10027}
\thanks[jnb]{jnb@ias.edu}\thanks[kamion]{kamion@phys.columbia.edu}

\begin{abstract}
In a series of recent papers, Ivanov et al. and Oberhummer et al. have
calculated the rate for the $p~+~p \rightarrow d~+~ e^+~+~ \nu_e$ 
reaction with a zero-range four-fermion effective interaction
and find a result 2.9 times higher than the
standard value calculated from non-relativistic potential theory.  
Their procedure is shown  to give a wrong answer because their assumed
interaction disagrees with low-energy $pp$ 
scattering data. 
\end{abstract}

\begin{keyword}
 Proton-proton reaction, Solar Fusion Reactions, Solar Neutrinos
\PACS{11.10.Ef,13.75.Cs,14.20.Dh\\
Preprint CU-TP-849, CAL-640, IASSNS-AST-97/45} 
\end{keyword}
\end{frontmatter}

In a series of recent papers, Ivanov et al. \cite{ivanov} and Oberhummer
et al. \cite{oberhummer} have calculated the rate for
the $p~+~p \rightarrow d~+~ e^+~+~ \nu_e$ 
reaction with a simplified  model for the
deuteron, and find a result 2.9 times higher than the current best
estimate, which is generally believed to 
be uncertain by  only a few percent \cite{kb,review}.  If
correct, the Ivanov et al. result would have
important implications for
stellar-evolution theory and for solar-neutrino research,
a fact that has been stressed in several recent articles  by 
Oberhummer et al. \cite{oberhummer}.

The calculation of Ivanov et al.
contradicts the  results obtained by  a long history of previous
researchers
(see, e.g., \cite{bc,salpeter,bm,kb}) beginning with Bethe and
Critchfield in 1938. All previous authors
find values for the $p~+~p \rightarrow d~+~ e^+~+~ \nu_e$ 
cross section which
agree with each other to within a few percent after accounting for 
small differences in the
experimentally determined input parameters. 
We show here that the result of Ivanov et al. is incorrect because
their assumed nuclear interactions do not fit the 
experimental data from low-energy $pp$ scattering.

We begin by recalling that 
low-energy $pp$ scattering experiments determine 
the $pp$ scattering length,
$a_p=-7.8196(26)$ fm, and effective range, $\rho_p=2.790(14)$ fm
\cite{bergervoet}.  

For the $pp$ interaction,
Ivanov  et al. postulate  an effective Lagrangian,
\begin{equation}
     {\cal L}_{\rm eff}^{pp} = - { g_{\pi N N}^2 C^2(v) \over 2 m_\pi^2}
     [\bar p \gamma_5 p][\bar p \gamma_5 p],
\label{lagrangian}
\end{equation}
where $g_{\pi N N}$ is a pion-nucleon-nucleon coupling, and $C(v)$ 
is the standard Gamow factor 
introduced in Eq.~(\ref{lagrangian}) 
on an {\it ad hoc} basis to partially describe
the Coulomb repulsion, with $v$ being the relative velocity in
the two-proton system.  At low energies,
Eq.~(\ref{lagrangian}) 
gives rise to a delta-function potential between the
two protons.  This delta function has an effective range $\rho_p=0$,
which is incorrect.  
We show below that this erroneous  effective range introduces an 
order-unity
error in the cross section for the 
$p~+~p \rightarrow d~+~ e^+~+~ \nu_e$ reaction and, by itself,
invalidates the calculation of Ivanov et al.

The scattering length does not appear directly in the effective
Lagrangian, Eq.~(\ref{lagrangian}), but is introduced in the one-loop,
long-wavelength approximation that Ivanov et al. use in their
calculations. 
They chose a 
value $a_p = -17.1$ fm \cite{nagels}, which has been corrected for
electromagnetic effects \cite{henley}.  
In the standard calculations \cite{kb,bc,salpeter,bm},
the Coulomb repulsion is included explicitly
in the proton-proton interaction and the nuclear interaction is 
then fit self-consistently to
the empirically determined scattering length, $-7.8196$ fm.  
Either choice of $a_p$ is an approximation using the Ivanov et
al. procedure since they only remove the asymptotic part of the 
Coulomb correction, $C(v)$.  Their treatment of the proton-proton
Coulomb repulsion 
introduces an additional 
error in their final result which is difficult to estimate.

How do we know that the calculated rate for the $pp$ reaction cannot
be wrong by a large factor?
Salpeter \cite{salpeter}, and later Bahcall and May \cite{bm},
showed that the rate for the $p~+~p\rightarrow d~+~ e^+~+~ \nu_e$
reaction can be determined  accurately given the 
measured $pp$ scattering
length and effective range, the  deuteron binding wave number
$\gamma$, the asymptotic $D$- to $S$-state ratio $\eta_d$, and the 
effective
range $\rho_d$.  This ``effective-range approximation''
makes no assumptions
about the details of the nuclear interactions, except that they must 
yield the measured values for these quantities. The success of the
effective-range  approximation 
relies on the fact that the matrix element for the reaction is
proportional to the overlap of the initial $pp$ wave function and the
final deuteron wave function and that most of the overlap comes
{}from radii large compared with the range of the 
nuclear forces.  The deuteron wave
function is constrained by the measured static deuteron parameters, and
the $pp$ wave function is constrained by low-energy $pp$-scattering data.
The effective-range approximation has been shown to agree to a few
percent with  numerical calculations of the matrix element for a wide
variety of assumed nuclear interactions \cite{kb,bm}.

We can use the effective-range  approximation to illustrate the effect of
invalid input data on the $p~+~p \rightarrow d~+~ e^+~+~ \nu_e$ 
cross section.  If, as in the Ivanov et al. Lagrangian, 
the $pp$ effective range were zero (with all other parameters fixed at
the experimentally determined values), the cross section would be about
$1.5$ times higher than the canonical value.  If we additionally assumed
the scattering length were twice the measured value, 
then the cross section would be 3
times larger than the standard result.
This numerical exercise shows that order-unity errors in $a_p$ or
$\rho_p$ will lead to order-unity errors in the $pp$ reaction rate.

In conclusion, the discrepant results of Ivanov et al.
are due (at least in part) to the fact that their assumed
$pp$ interaction disagrees with low energy scattering data.
Our previous results for a broad range of nuclear
interactions show  \cite{kb,bm} that any calculation which is
consistent with the measured low-energy scattering of the $pp$ system
and the measured properties of the deuteron will yield a cross section
that differs from the current best-estimate by no more than a
few percent.\footnote{We do not discuss the
difficult questions associated with the general procedure adopted by 
Ivanov et al. \cite{ivanov}, in which nucleons and light nuclei are 
treated as fundamental particles in a relativistic field theory.
This program is complicated because there is no small parameter that
makes perturbative calculations reliable and because the effective
theory should be derived from QCD.}\footnote{Degl'Innocenti et
al. have also recently pointed out that an enhancement of the $pp$
reaction rate by a factor of 2.9 would be in gross disagreement
with helioseismological data \cite{degl}.}

\bigskip
We are grateful to C. Callan, E. Henley, P. Krastev, C. Nappi, and 
S. Treiman for valuable
discussions.
MK was supported by the D.O.E. grant number DEFG02-92-ER 40699, NASA
NAG5-3091, and the Alfred P. Sloan Foundation.  JNB was supported
by NSF grant \#PHY95-13835.

\end{document}